\documentclass[aps,pre,amsmath,amssymb,twocolumn,showpacs,superscriptaddress]{revtex4}     
\usepackage{graphicx}

\begin{document}
\title{Crystal Nucleation in a Supercooled Liquid with Glassy Dynamics}

\author{Ivan Saika-Voivod}
\affiliation{Department of Physics and Physical Oceanography,
Memorial University of Newfoundland, St. John's, NL, A1B 3X7, Canada}

\author{Richard K. Bowles}
\affiliation{Department of Chemistry, University of Saskatchewan, Saskatoon, SK, 57N 5C9, Canada}

\author{Peter H. Poole}
\affiliation{Department of Physics, St. Francis Xavier University,
Antigonish, NS, B2G 2W5, Canada}

\begin{abstract}
In simulations of supercooled, high-density liquid silica we study a range of temperature $T$ in which we find both crystal nucleation, as well as the characteristic dynamics of a glass forming liquid, including a breakdown of the Stokes-Einstein relation.  We find that the liquid cannot be observed below a homogeneous nucleation limit (HNL) at which the liquid crystallizes faster than it can equilibrate.  We show that the HNL would 
occur at lower $T$, and perhaps not at all, if the Stokes-Einstein relation were obeyed, and hence that glassy dynamics plays a central role in setting a crystallization limit on the liquid state in this case.  We also explore the relation of the HNL to the Kauzmann temperature, and test for spinodal-like effects near the HNL.
\end{abstract}

\date{October 19, 2009}
\maketitle

A long-standing question regarding the liquid state concerns its ultimate fate when supercooled below the equilibrium freezing temperature~\cite{pablo}.  In his seminal 1948 work on the glass transition, Kauzmann pointed out that many supercooled liquids are headed for an ``entropy catastrophe'' as the temperature $T$ decreases~\cite{kauzmann}.  That is, the liquid entropy decreases so rapidly with $T$ that unless crystallization or the glass transition intervenes, the entropy will become negative below a finite temperature $T_K$.  Since 1948, two major scenarios have emerged for avoiding the entropy catastrophe based on our understanding of glasses:  that the liquid terminates at $T_K$ in an ``ideal glass" state; and that a ``fragile-to-strong" crossover occurs in which the entropy changes its $T$-dependence so as to remain non-zero for all $T>0$~\cite{pablo,tanaka}.

Yet it is often crystallization, rather than the glass transition, that terminates the liquid on cooling prior to $T_K$.  In experiments, a homogeneous nucleation limit (HNL) is often encountered below which crystallization is, in practice, unavoidable~\cite{pablo}.  In theoretical and simulation studies, a kinetically-defined HNL has been identified as the $T$ at which the time for crystal nucleation $\tau_n$ becomes comparable to the structural relaxation time, $\tau_\alpha$~\cite{kauzmann,kiselev,tanaka,cavagna}.  Below this HNL, the supercooled liquid ceases to be observable because it nucleates before it can equilibrate.  Other recent simulation studies report the occurrence of a thermodynamically-defined spinodal-like limit for crystal nucleation, at which the free energy barrier to nucleation $\Delta G^\ast$ decreases to zero\cite{wolde,parrinello,wang}.

Kauzmann himself proposed that crystallization becomes inevitable in supercooled liquids prior to $T_K$~\cite{kauzmann}, and recently Tanaka has reexamined this scenario, through an analysis of both classical nucleation theory (CNT) and the behavior of glass forming liquids~\cite{tanaka}. While CNT forms the basis of much of our understanding of nucleation, by itself it seems to predict no limit on supercooling~\cite{pablo}.  In CNT the nucleation time is given by $\tau_n=K \exp (\Delta G^\ast/RT)$, where $R$ is the gas constant, and the kinetic prefactor $K$ contains a factor of $D^{-1}$, the inverse of the liquid diffusion coefficient.  Assuming that $D^{-1}$ is proportional to $\tau_\alpha$, a system obeying CNT will always satisfy $\tau_n>\tau_\alpha$, i.e. the equilibrium liquid will be observable prior to nucleation at all $T$. 

In this context, Tanaka pointed out that a liquid can obey CNT and exhibit a HNL~\cite{hnl} if there is a breakdown of the Stokes-Einstein (SE) relation~\cite{tanaka}.  The SE relation asserts that $D \eta/T$, where $\eta$ is the viscosity, is independent of $T$.  Here we use $\tau_\alpha$ as a proxy for $\eta$, since both quantify collective structural relaxation.  The violation of the SE relation is a ubiquitous feature of supercooled liquids, in which $D\tau_\alpha/T$ is found to grow rapidly as $T$ decreases, demonstrating that the relaxation time associated with diffusion (a local process) increases faster than $\tau_\alpha$ (a global process).  Since in CNT $\tau_n$ is controlled by $D$, and not by $\tau_\alpha$, SE breakdown makes it possible for $\tau_n$ and $\tau_\alpha$ to become comparable at sufficiently low $T$, inducing a HNL.

Tanaka's analysis is significant because it predicts that the physics of a glass forming liquid (SE breakdown) is crucial to the origins of the phenomenon (unavoidable crystallization at the HNL) by which the entropy catastrophe is avoided.  Tanaka showed that experimental data on a metallic liquid alloy is consistent with his interpretation~\cite{tanaka}.  Cavagna and coworkers have come to similar conclusions by incorporating SE breakdown into CNT through viscoelastic effects~\cite{cavagna}.  In this Letter we use computer simulations to identify a HNL in a deeply supercooled liquid, and show that the nature and location of this limit is indeed strongly influenced by the presence of glassy dynamics.

\begin{figure}[t]
\centerline{\includegraphics[scale=0.45]{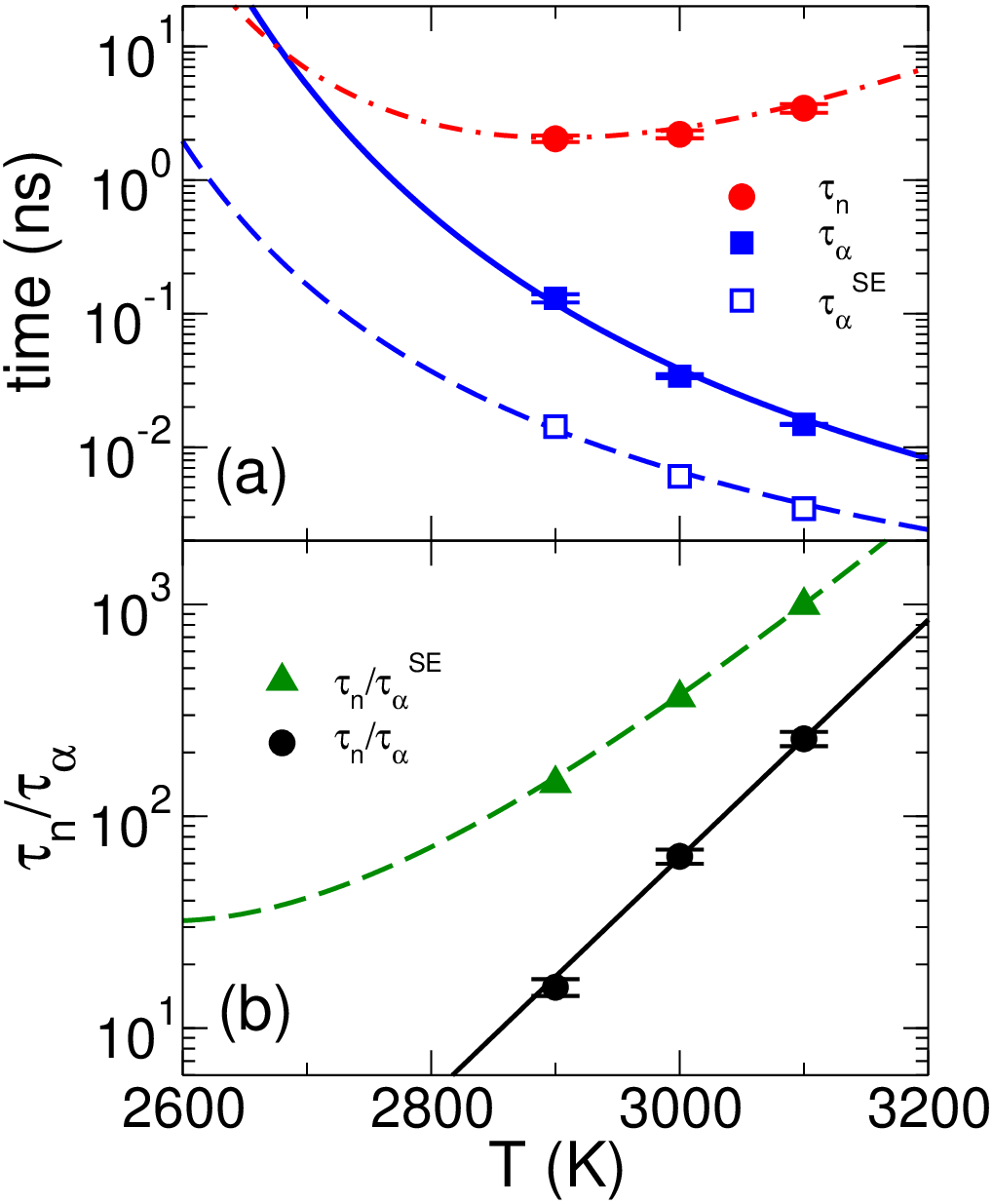}}
\caption{(a) Behavior of $\tau_\alpha$, $\tau_n$ and $\tau_\alpha^{\rm SE}$ 
as a function of $T$.  The model functions $\tau_\alpha^{\rm VFT}$ (solid line), $\tau_\alpha^{\rm VFT}/R$ (dashed line) and $\cal{E}\tau_\alpha^{\rm VFT}$ (dot-dashed line) are also shown.
(b) $T$-dependence of $\tau_n/\tau_\alpha$ and $\tau_n/\tau_\alpha^{\rm SE}$. The model functions $\cal{E}$ (solid line) and
${\cal{E}}R$ (dashed line) are also shown.
}
\label{times}
\end{figure}

\begin{figure}[t]
\centerline{\includegraphics[scale=0.3]{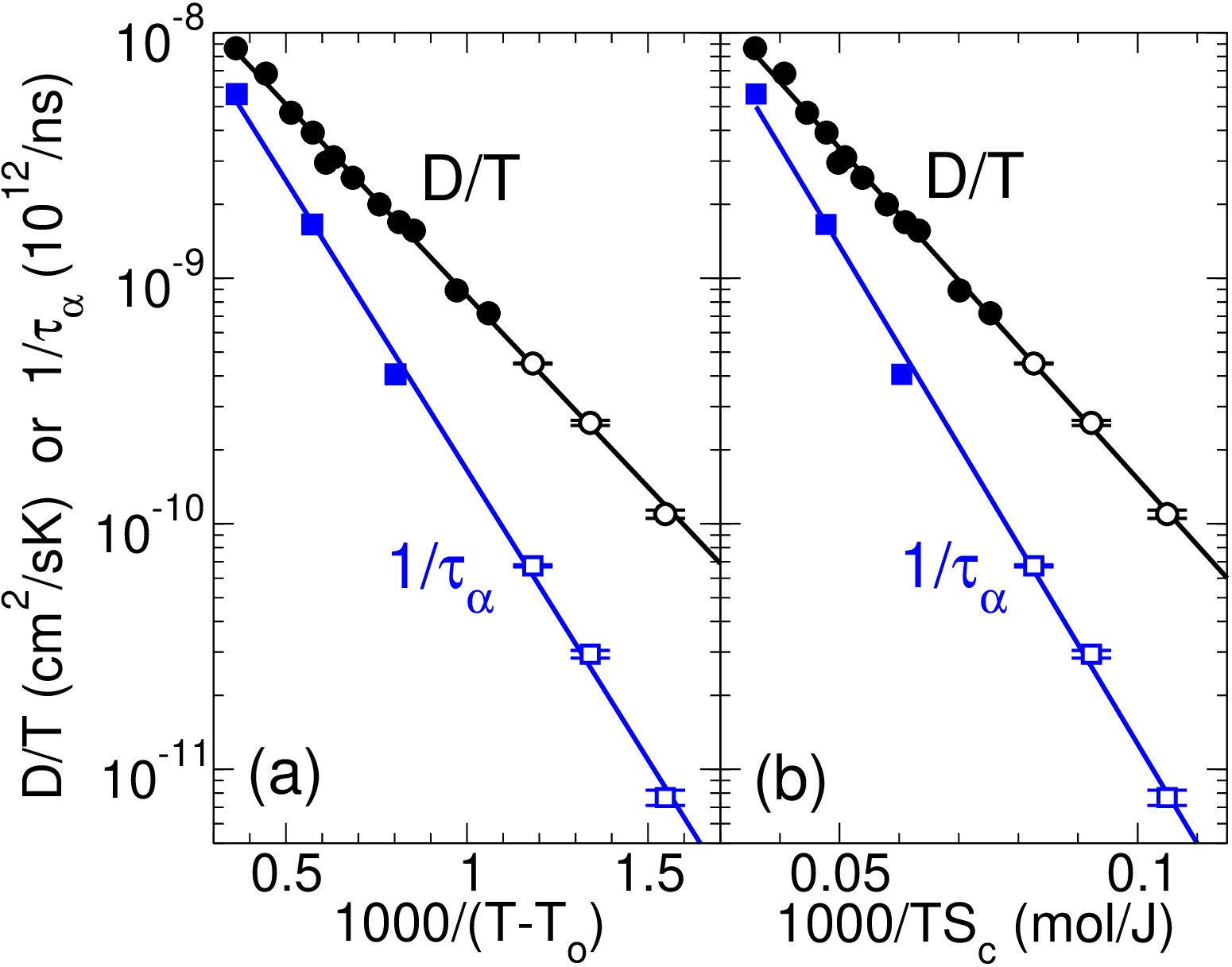}}
\caption{(a) VFT and (b) AG plots of $D/T$ (circles) and $1/\tau_\alpha$ (squares).  Filled symbols are data generated for $T\geq3200$~K from single simulation runs.  Open symbols are data found using the 20 longest runs from the ensemble of 200 conducted at each $T<3200$~K.  Fits of a straight line to each data set are also shown.  In (a) $T_o=2254$~K.  In (b) the data for $S_c$ from Fig.~\protect{\ref{se}}(a) is used to evaluate $1/TS_c$.}
\label{ag}
\end{figure}

Our results are based on molecular dynamics simulations of the BKS model of silica (SiO$_2$)~\cite{bks}.  Our system consists of 444 Si atoms and 888 O atoms at fixed volume, with simulation parameters the same as in Refs.~\cite{SPS,SPB}.  This model liquid is a well-studied glass-former~\cite{bks-work}, and yet also crystallizes to stishovite on time scales accessible to simulation for $T<3200$~K at density $4.38$~g/cm$^3$~\cite{SPB}.   Under these conditions the liquid is deeply supercooled; at this liquid density the liquid-stishovite coexistence temperature is greater than 6000~K~\cite{SPB}.  

We quantify the dynamical properties of the liquid by evaluating $D$ and $\tau_\alpha$ at density $4.38$~g/cm$^3$ for several $T$ from $5000$ to $2900$~K.  
$D$ is evaluated from the mean square displacement, while $\tau_\alpha$ is defined as the time constant in a fit of a stretched exponential $\exp[-(t/\tau_\alpha)^\beta]$ to the decay of the intermediate scattering function at a wavenumber corresponding to the first peak of the structure factor.  Both quantities are computed for the Si atoms only.  For each $T \geq 3200$~K, these properties are evaluated from a microcanical simulation run, starting from a well-equilibrated initial configuration.

For $T<3200$~K, crystallization occurs spontaneously on our computational time scale.  In this regime we seek to evaluate $D$ and $\tau_\alpha$ for the liquid as well as to quantify the crystal nucleation kinetics.  To this end, for each $T<3200$~K, we conduct 200 runs initiated from distinct configurations equilibrated at $5000$~K, and quench the system by applying the Berendsen thermostat~\cite{beren} (with a time constant of 1~ps) to decrease $T$ to the target value.  Each run continues until it crystallizes, as detected by a significant drop in the potential energy.  The nucleation time for a run is taken as the latest time at which the system contains no crystalline particles~\cite{fn-cryst}.  These times are averaged over the 200 runs to give the mean nucleation time $\tau_n$; uncertainties are computed as the standard deviation of the mean.  

To estimate $D$ and $\tau_\alpha$ for each $T<3200$~K, we identify those 20 of the 200 runs that remain longest in the liquid state before crystallizing.  At three $T$ (3100, 3000 and 2900~K) we find that the transient behavior associated with the quenching procedure is completely removed by discarding the initial 2~ns of each run.  The liquid state properties (e.g. energy, pressure) are stationary thereafter, and we find $D$ and $\tau_\alpha$ for each run from this stationary time series.  Averaging over the 20 runs, we compute the mean value of $D$ and $\tau_\alpha$, and their respective uncertainties as the standard deviation of the mean.  In this work, we restrict our attention to those $T$ at which we can confirm that liquid equilibrium is established on a time scale much less than the nucleation time (i.e. $\tau_\alpha<<\tau_n$) in order to ensure that nucleation events do not interfere with the accurate evaluation of liquid properties.

It is important to note that $\tau_n$ reported here is the {\it system} nucleation time, which depends on the system volume $V$ as $\tau_n=(JV)^{-1}$ where $J$ is the nucleation rate per unit volume~\cite{pablo}.   Hence $\tau_n$ for larger systems than ours will be smaller by a factor of $V_o/V$, where $V_o$ is our system volume.  In simulations, as in experiments, small systems are often exploited in order to increase the nucleation time and thus allow examination of deeply supercooled liquid states.  As shown below, our system size is small enough to allow us to reach a range of $T$ in which both nucleation and glassy dynamics occur; and yet large enough to easily accommodate crystal nuclei of critical size.

\begin{figure}[t]
\centerline{\includegraphics[scale=0.34]{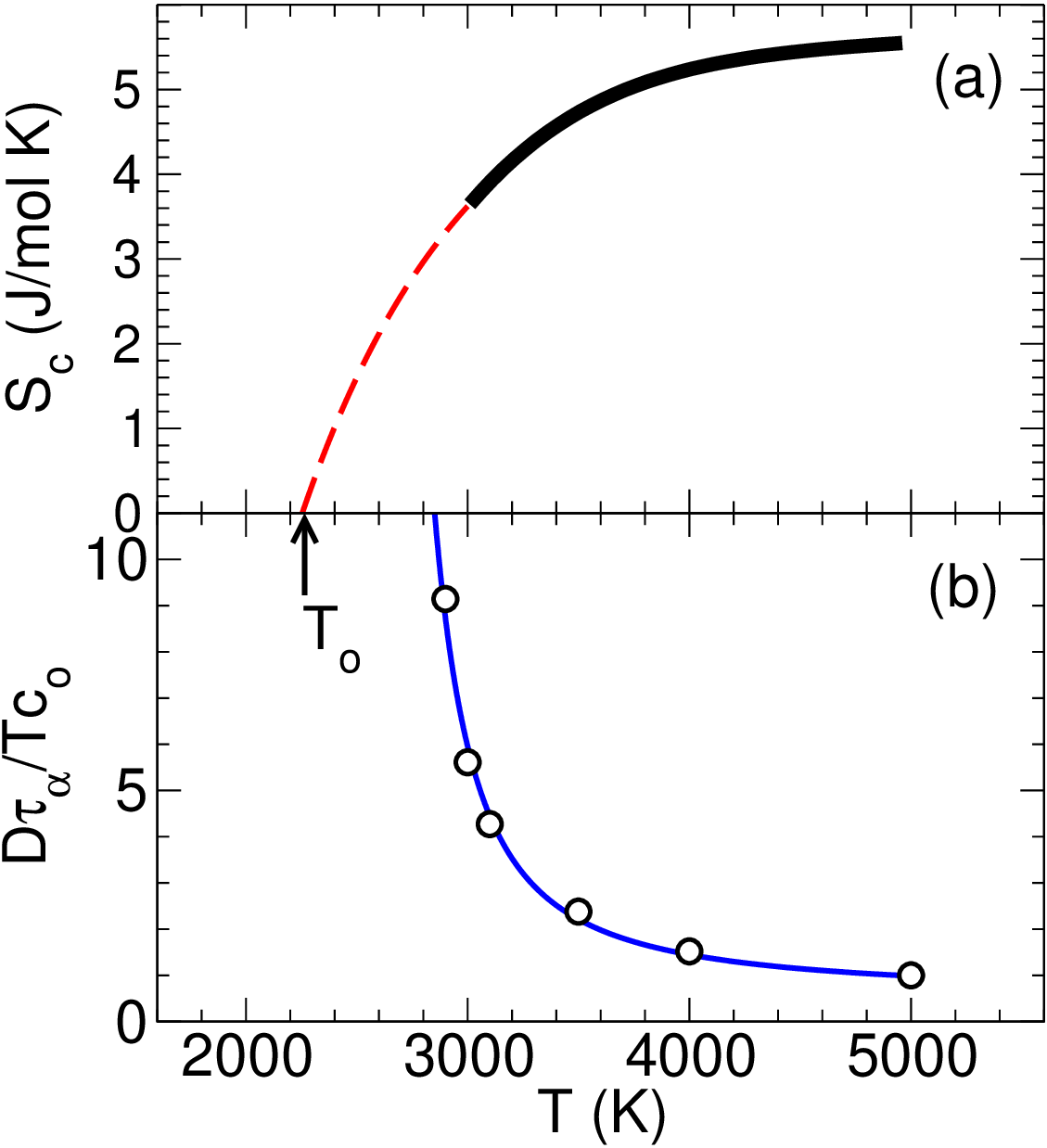}}
\caption{(a) $S_c$ as a function of $T$ (thick solid line).  
The dashed line is an extrapolation based on the finding in Fig.~\protect{\ref{ag}} that both the VFT and AG relations are satisfied.  That is, we solve $A\exp[B/(T-T_o)]=C\exp[E/(TS_c)]$ for $S_c$, where the fitting parameters $A$, $B$, $T_o$, $C$ and $E$ are taken from the fits to $D/T$ in Fig.~\protect{\ref{ag}}.  This curve models the case where $S_c$ approaches an ideal glass transition at $T_K=T_o$.  (b) $T$ dependence of the SE ratio $D\tau_\alpha/T$, normalized by $c_o$, the value of $D\tau_\alpha/T$ at $5000$~K.  The solid line is the model function $R$.
}
\label{se}
\end{figure}

In Fig.~\ref{times}(a) the $T$-dependence of $\tau_n$ is compared to that of $\tau_\alpha$.  Previous studies indicate that liquid equilibrium is established on a time scale of between 10 and 20 times $\tau_\alpha$~\cite{cavagna1,tanaka-mro}.  At $T=3100$, 3000 and 2900~K, we find that $\tau_n$ remains greater than an order of magnitude larger than $\tau_\alpha$, indicating that liquid equilibrium is well established prior to crystal nucleation.  However, the gap between $\tau_n$ and $\tau_\alpha$ is closing rapidly as $T$ decreases.  Fig.~\ref{times}(b) shows that the ratio $\tau_n/\tau_\alpha$ is decreasing with $T$ in a manner suggesting that this liquid system reaches a HNL in the vicinity of $2800$~K, where nucleation will on average occur faster than equilibrium measurements of liquid properties can be made.  Consistent with this behavior, we have attempted simulations at 2800~K, but find that a significant fraction of the 200 runs nucleate during the initial transient associated with the quenching procedure, complicating the evaluation of $\tau_n$, as well as of $D$ and $\tau_\alpha$.  Empirically, we also note that in the $T$ range studied here $\tau_n/\tau_\alpha$ fits well to an exponential function (referred to below as $\cal{E}$), as shown in Fig.~\ref{times}(b).

On approach to the HNL, we find that the liquid behaves as a fragile glass former, in that $D/T$ and $\tau_\alpha$ are modeled well by the Vogel-Fulcher-Tammann (VFT) expression~\cite{VFT}, $A\exp[B/(T-T_o)]$, where $A$, $B$ and $T_o$ are fitting parameters [Fig.~\ref{ag}(a)].  The dynamical divergence temperature $T_o$ serves as an estimate of $T_K$.  We obtain $T_o=2270$ and $2237$~K for $D/T$ and $\tau_\alpha$ respectively, and in Fig.~\ref{ag}(a) we set $T_o$ to the mean of these values, $2254$~K.  
The VFT fit to $\tau_\alpha$, denoted $\tau_\alpha^{\rm VFT}$, is shown in Fig.~\ref{times}(a).

The fragile nature of the liquid is also demonstrated in the $T$-dependence of the configurational entropy $S_c$ [Fig.~\ref{se}(a)].  $S_c$ quantifies the entropy associated with the number of distinct basins of the potential energy landscape explored by the liquid at a given $T$.  We evaluate $S_c$ via an analysis of the inherent structure energy of the liquid, as described in detail in Ref.~\cite{SPS}.  We find that $S_c$ decreases rapidly as $T$ decreases, characteristic of a fragile glass former.  Using $S_c$ we also find, in common with many glass forming liquids, that $D/T$ and $\tau_\alpha$ satisfy the Adam-Gibbs (AG) expression~\cite{ag}, $C\exp[E/(TS_c)]$, where $C$ and $E$ are fitting parameters [Fig.~\ref{ag}(b)].  The fact that $D/T$ and $\tau_\alpha$ conform to both the VFT and AG relations at all $T$ in Fig.~\ref{ag} is a validation of our method for finding the equilibrium liquid behavior in the low $T$ range, where nucleation also occurs.

A further signature of glassy dynamics is shown in Fig.~\ref{se}(b), which gives the SE ratio $D\tau_\alpha/T$ normalized by its high-$T$ value $c_o$.  As $T$ decreases, the SE relation breaks down, and at the lowest $T$ the characteristic time scale for structural relaxation is nearly 10 times larger than that 
associated with the diffusion process, compared to high $T$.  
We use the VFT fits to $D/T$ and $\tau_\alpha$ to evaluate a model function (denoted $R$) for $D\tau_\alpha/Tc_o$, shown in Fig.~\ref{se}(b).

\begin{figure}[t]
\centerline{\includegraphics[scale=0.28]{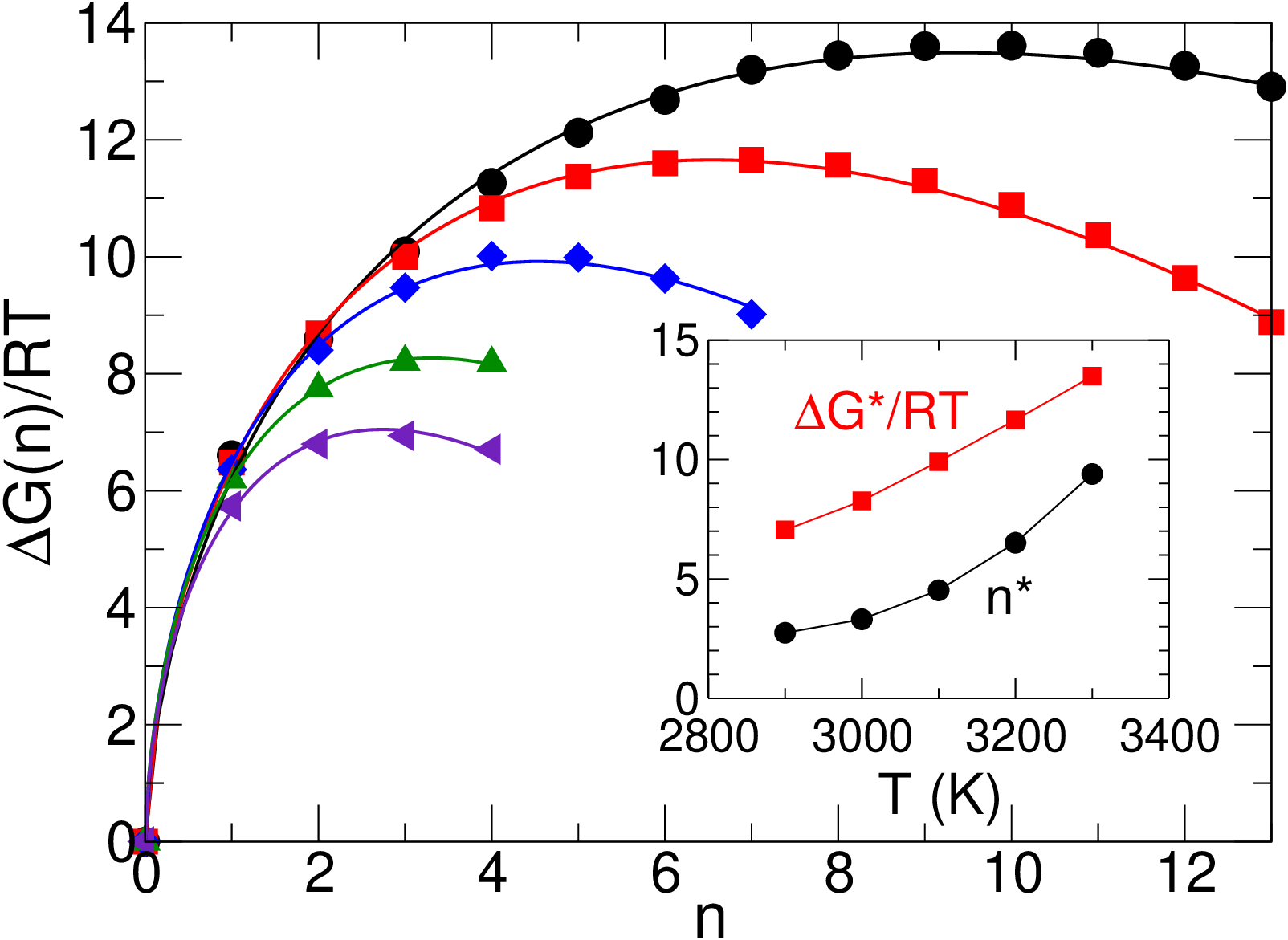}}
\caption{$\Delta G(n)/RT$ as function of $n$ for $T=2900$, 3000, 3100, 3200 and 3300~K (data sets from bottom to top).  Solid curves are fits to the CNT prediction $\Delta G(n)/RT=-an+bn^{2/3}$, where $a$ and $b$ are fitting parameters.  Inset: $\Delta G^\ast/RT$ and $n^\ast$ as a function of $T$.}
\label{profiles}
\end{figure}

Next we seek to quantify the role played by SE breakdown in the occurrence of the HNL.  To do so, we estimate what $\tau_\alpha$ would be if the SE relation were obeyed.  That is, we define $\tau_\alpha^{\rm SE}$ as the value of $\tau_\alpha$ computed using the data for $D$ via the SE relation, i.e. $\tau_\alpha^{\rm SE}=c_oT/D$.  If we assume that $\tau_n$ depends on $D$, but not on $\tau_\alpha$, as predicted by CNT, then $\tau_n$ can be compared to $\tau_\alpha^{\rm SE}$ to test for the occurrence of a HNL if the SE relation were obeyed.  

We find that $\tau_\alpha^{\rm SE}$ is as much as an order of magnitude smaller than $\tau_\alpha$, and that $\tau_n$ and $\tau_\alpha^{\rm SE}$ are more widely separated, and converging less rapidly, than $\tau_n$ and $\tau_\alpha$ [Fig.~\ref{times}(a)].  In Fig.~\ref{times}(b) we plot the ratio $\tau_n/\tau_\alpha^{\rm SE}$.  Noting that $\tau_n/\tau_\alpha^{\rm SE}=(\tau_n/\tau_\alpha)(D\tau_\alpha/Tc_o)$, we model $\tau_n/\tau_\alpha^{\rm SE}$ as the product of the fitting functions $\cal{E}$ and $R$.  In Fig.~\ref{times}(a), we model $\tau_\alpha^{\rm SE}$ as $\tau_\alpha^{\rm VFT}/R$, and $\tau_n$ as $\cal{E}\tau_\alpha^{\rm VFT}$, to show the behavior consistent with the modelling presented in Fig.~\ref{times}(b).

Comparison of $\tau_n/\tau_\alpha$ and $\tau_n/\tau_\alpha^{\rm SE}$ in Fig.~\ref{times}(b) suggests that if the SE relation were obeyed, the HNL would be shifted to lower $T$ by at least several hundred degrees~K.  Since $\tau_n/\tau_\alpha$ and $\tau_n/\tau_\alpha^{\rm SE}$ differ by a factor of $D\tau_\alpha/Tc_o$, which appears to diverge as $T\to T_o$, the gap between $\tau_n/\tau_\alpha$ and $\tau_n/\tau_\alpha^{\rm SE}$ will grow rapidly as $T$ decreases.  So long as $\tau_n/\tau_\alpha$ decreases with $T$ more slowly than $D\tau_\alpha/Tc_o$ increases, it is possible for the liquid to remain observable (i.e. satisfy $\tau_\alpha<<\tau_n$).  For example, if the exponential fit to $\tau_n/\tau_\alpha$ holds for $T<2900$~K, our data allow the possibility that a liquid obeying the SE relation remains observable approaching $T_o$.

In sum the above results illustrate a case in which SE breakdown plays a key role in setting a supercooling limit on the liquid state due to crystallization.  This interplay of glassy phenomena and crystal formation is a realization of Kauzmann's original proposal for avoiding the entropy catastrophe at $T_K$, and is entirely consistent with Tanaka's recent analysis~\cite{tanaka}.  In our system, $T_K$ cannot be approached because of a HNL at approximately 2800~K.  Yet, in the absence of glassy dynamics (in the form of SE breakdown), lower $T$ would be accessible, including perhaps the region near $T_K$.  

We note that the effect of system size on $\tau_n$ explained above will not qualitatively change our conclusions.  Decreasing $V$, for example by a factor of 10 (from 444 to 44 molecules), will increase $\tau_n$ by a factor of 10 and lower the HNL by approximately 200~K.  This would place the HNL at approximately 2600~K, still well above $T_o=2254$~K.  It would be difficult to justify any further reduction in $V$ that would not introduce significant, unphysical finite-size effects.  The HNL studied here is thus located near the lowest possible $T$ at which such a limit can be observed in this system, and yet is always well above $T_o$ for all $V$.

Finally, we also examine the thermodynamic aspects of the nucleation process, to test if the kinetically-defined HNL is related to a spinodal-like thermodynamic limit.    We show in Fig.~\ref{profiles} $\Delta G(n)$, the work of formation of crystalline clusters of size $n$ for several $T$.  Our procedure for computing $\Delta G(n)$ is the same as that used in Ref.~\cite{SPB}.  The number of molecules in the critical nucleus $n^\ast$, as well as $\Delta G^\ast$ are both decreasing as a function of $T$ (inset, Fig.~\ref{profiles}).  At the same time, both quantities remain finite in the range of $T$ studied here, and the shape of $\Delta G(n)$ remains consistent with CNT.   Further, in nucleation influenced by a spinodal, it is expected that the critical nucleus becomes ramified and/or that the nucleation process involves a coalescence of distinct crystalline clusters~\cite{parrinello,wang}.  We find no indication of such behavior in our system.  At almost all times during the nucleation process, the system contains at most one compact crystalline cluster that is of critical size or greater.   

Nucleation thus remains a localized, activated process as $T$ decreases toward the HNL, and we do not find evidence that spinodal-like phenomena influence the nature of the liquid state in this range of $T$.  This result is consistent with a recent study of nucleation in the Ising model in which nucleation remained classical even when the nucleation barrier was less than $8RT$~\cite{new-prl}.  These findings reinforce the strongly kinetic (rather than thermodynamic) character of the HNL found here, and the key role played by glassy dynamics on approach to this limit.

We thank ACEnet and SHARCNET for providing computational resources, and NSERC for financial support.  PHP thanks the CRC program for support.


\begin{thebibliography}{999}

\bibitem{pablo}
P.G. Debenedetti, {\it Metastable Liquids. Concepts and Principles} (Princeton University Press, Princeton, New Jersey, 1996).

\bibitem{kauzmann}  W. Kauzmann, Chem. Rev. (Washington, D.C.) {\bf 43}, 219 (1948).

\bibitem{tanaka} H. Tanaka, Phys. Rev. E {\bf 68}, 011505 (2003).

\bibitem{kiselev} S.B. Kiselev and J.F. Ely, Physica A {\bf 299}, 357 (2001).

\bibitem{cavagna} A. Cavagna, A. Attanasi and J. Lorenzana, Phys. Rev. Lett. {\bf 95}, 115702 (2005).   

\bibitem{wolde} P.R. ten Wolde, M.J. Montero and D. Frenkel, J. Chem. Phys. {\bf 104}, 9932 (1996).

\bibitem{parrinello} F. Trudu, D. Donadio and M. Parrinello, Phys. Rev. Lett. {\bf 97}, 105701 (2006).

\bibitem{wang} H. Wang, H. Gould and W. Klein, Phys. Rev. E {\bf 76}, 031604 (2007).

\bibitem{hnl}    In Ref.{\protect{\cite{tanaka}}}, Tanaka uses the term ``lower metastable limit" to refer to a kinetically-defined HNL.  

\bibitem{bks} B.W.H. van Beest, G.J. Kramer and R.A. van Santen, Phys. Rev. Lett. {\bf 64}, 1955 (1990).

\bibitem{SPB}  I. Saika-Voivod, P.H. Poole and R.K. Bowles, J. Chem. Phys. {\bf 124}, 224709 (2006).

\bibitem{SPS} I. Saika-Voivod, P.H. Poole and F. Sciortino, Nature (London) {\bf 412}, 514 (2001); I. Saika-Voivod, F. Sciortino and P. H. Poole, Phys. Rev. E {\bf 69}, 041503 (2004).

\bibitem{bks-work} J. Horbach and W. Kob, Phys. Rev. B {\bf 60}, 3169 (1999); I. Saika-Voivod, F. Sciortino, and P.H. Poole, Phys. Rev. E {\bf 63}, 011202 (2000); I. Saika-Voivod, et al., Phys. Rev. E {\bf 70}, 061507 (2004).

\bibitem{beren} H.J.C. Berendsen, et al., J. Chem. Phys. {\bf 81}, 3684 (1984).

\bibitem{fn-cryst}  To define crystalline particles we focus on Si atoms only and use the approach described in Ref.~\protect{\cite{wolde}}.
See also Ref.~\protect{\cite{SPB}} for details.

\bibitem{cavagna1} A. Cavagna, I. Giardina and T.S. Grigera, J. Chem. Phys {\bf 118}, 6974 (2003).

\bibitem{tanaka-mro} H. Shintani and H. Tanaka, Nature Physics {\bf 2}, 200 (2006).

\bibitem{VFT} H. Vogel, Phys. Zeit. {\bf 22}, 645 (1921); G.S. Fulcher, J. Am. Ceram. Soc. {\bf 8}, 339 (1925); G. Tammann, J. Soc. Glass Technol. {\bf 9}, 166 (1925).

\bibitem{ag} G. Adam and J.H. Gibbs, J. Chem. Phys. {\bf 43}, 139 (1965). 

\bibitem{new-prl} L. Maibaum, Phys. Rev. Lett. {\bf 101}, 256102 (2008).

\end{thebibliography}
\end{document}